\documentclass[twocolumn]{aastex63}

\newcommand\teff{\mbox{$T_\mathrm{eff}$}}
\usepackage{color}

\begin{document}

\title{Ross 19B: An Extremely Cold Companion Discovered via the Backyard Worlds: Planet 9 Citizen Science Project}

\correspondingauthor{Adam C. Schneider}
\email{aschneid10@gmail.com}

\author[0000-0002-6294-5937]{Adam C. Schneider}
\affil{United States Naval Observatory, Flagstaff Station, 10391 West Naval Observatory Rd., Flagstaff, AZ 86005, USA}
\affil{Department of Physics and Astronomy, George Mason University, MS3F3, 4400 University Drive, Fairfax, VA 22030, USA}

\author[0000-0002-1125-7384]{Aaron M. Meisner}
\affil{NSF's National Optical-Infrared Astronomy Research Laboratory, 950 N. Cherry Ave., Tucson, AZ 85719, USA}

\author[0000-0002-2592-9612]{Jonathan Gagn\'e}
\affil{Plan\'etarium Rio Tinto Alcan, Espace pour la Vie, 4801 ave. Pierre-de Coubertin, Montr\'eal, QC H1V~3V4, Canada}
\affil{Institute for Research on Exoplanets, Universit\'e de Montr\'eal, 2900 Boulevard \'Edouard-Montpetit Montr\'eal, QC H3T~1J4, Canada}

\author[0000-0001-6251-0573]{Jacqueline K. Faherty}
\affil{Department of Astrophysics, American Museum of Natural History, Central Park West at 79th St., New York, NY 10024, USA}

\author[0000-0001-7519-1700]{Federico Marocco}
\affil{IPAC, Mail Code 100-22, Caltech, 1200 E. California Blvd., Pasadena, CA 91125, USA}

\author[0000-0002-6523-9536]{Adam J. Burgasser}
\affil{Center for Astrophysics and Space Science, University of California San Diego, La Jolla, CA 92093, USA}

\author[0000-0003-4269-260X]{J. Davy Kirkpatrick}
\affil{IPAC, Mail Code 100-22, Caltech, 1200 E. California Blvd., Pasadena, CA 91125, USA}

\author[0000-0002-2387-5489]{Marc J. Kuchner}
\affil{Exoplanets and Stellar Astrophysics Laboratory, NASA Goddard Space Flight Center, 8800 Greenbelt Road, Greenbelt, MD 20771, USA}

\author[0000-0002-8960-4964]{L\'eopold Gramaize}
\affil{Backyard Worlds: Planet 9, USA}

\author[0000-0003-4083-9962]{Austin Rothermich}
\affil{Backyard Worlds: Planet 9, USA}

\author[0000-0002-5253-0383]{Hunter Brooks}
\affil{Department of Physics and Astronomy, Northern Arizona University, Flagstaff, AZ
86011, USA}

\author{Frederick J. Vrba}
\affil{United States Naval Observatory, Flagstaff Station, 10391 West Naval Observatory Rd., Flagstaff, AZ 86005, USA}

\author[0000-0001-8170-7072]{Daniella Bardalez Gagliuffi}
\affil{Department of Astrophysics, American Museum of Natural History, Central Park West at 79th Street, NY 10024, USA}

\author[0000-0001-7896-5791]{Dan Caselden}
\affil{Backyard Worlds: Planet 9, USA}

\author[0000-0001-7780-3352]{Michael C. Cushing}
\affil{Ritter Astrophysical Research Center, Department of Physics and Astronomy, University of Toledo, 2801 W. Bancroft St., Toledo, OH 43606, USA}

\author{Christopher R. Gelino}
\affil{IPAC, Mail Code 100-22, Caltech, 1200 E. California Blvd., Pasadena, CA 91125, USA}

\author[0000-0002-2387-5489]{Michael R. Line}
\affil{School of Earth \& Space Exploration, Arizona State University, Tempe, AZ 85287, USA}

\author[0000-0003-2478-0120]{Sarah L. Casewell}
\affiliation{Department of Physics and Astronomy, University of Leicester, University Road, Leicester LE1 7RH, UK}

\author[0000-0002-1783-8817]{John H. Debes}
\affil{ESA for AURA, Space Telescope Science Institute, 3700 San Martin Drive, Baltimore, MD 21218, USA
}

\author[0000-0003-2094-9128]{Christian Aganze}
\affil{Center for Astrophysics and Space Science, University of California San Diego, La Jolla, CA 92093, USA}

\author{Andrew Ayala}
\affil{Department of Astrophysics, American Museum of Natural History, Central Park West at 79th St., New York, NY 10024, USA}

\author[0000-0003-0398-639X]{Roman Gerasimov}
\affil{Center for Astrophysics and Space Science, University of California San Diego, La Jolla, CA 92093, USA}

\author[0000-0003-4636-6676]{Eileen C. Gonzales}
\altaffiliation{51 Pegasi b Fellow}
\affil{Department of Astronomy and Carl Sagan Institute, Cornell University, 122 Sciences Drive, Ithaca, NY 14853, USA}

\author[0000-0002-5370-7494]{Chih-Chun Hsu}
\affil{Center for Astrophysics and Space Science, University of California San Diego, La Jolla, CA 92093, USA}

\author[0000-0003-2102-3159]{Rocio Kiman}
\affil{Department of Astrophysics, American Museum of Natural History, Central Park West at 79th St., New York, NY 10024, USA}
\affil{Department of Physics, Graduate Center, City University of New York, 365 5th Avenue, New York, NY 10016, USA}

\author{Mark Popinchalk}
\affil{Department of Astrophysics, American Museum of Natural History, Central Park West at 79th St., New York, NY 10024, USA}
\affil{Department of Physics, Graduate Center, City University of New York, 365 5th Avenue, New York, NY 10016, USA}

\author[0000-0002-9807-5435]{Christopher Theissen}
\altaffiliation{NASA Sagan Fellow}
\affil{Center for Astrophysics and Space Science, University of California San Diego, La Jolla, CA 92093, USA}

\author{The Backyard Worlds: Planet 9 Collaboration}

\begin{abstract}

Through the Backyard Worlds: Planet 9 citizen science project, we have identified a wide-separation ($\sim$10\arcmin, $\sim$9900 au projected) substellar companion to the nearby ($\sim$17.5 pc), mid-M dwarf Ross 19. We have developed a new formalism for determining chance alignment probabilities based on the BANYAN $\Sigma$ tool, and find a 100\% probability that this is a physically associated pair.  Through a detailed examination of Ross 19A, we find that the system is metal-poor ([Fe/H]=$-$0.40$\pm$0.12) with an age of 7.2$^{+3.8}_{-3.6}$ Gyr.  Combining new and existing photometry and astrometry, we find that Ross 19B is one of the coldest known wide-separation companions, with a spectral type on the T/Y boundary, an effective temperature of 500$^{+115}_{-100}$ K, and a mass in the range 15--40 $M_{\rm Jup}$.  This new, extremely cold benchmark companion is a compelling target for detailed characterization with future spectroscopic observations using facilities such as the {\it Hubble Space Telescope} or {\it James Webb Space Telescope}.  

\end{abstract}


\section{Introduction}
\label{sec:introduction}

Brown dwarfs and giant planets never settle onto a stable position in color-magnitude diagrams; instead, they radiatively cool along evolutionary tracks that are degenerate in mass, metallicity, luminosity, and age (e.g., \citealt{burrows2001}). Such evolution makes determining fundamental properties of brown dwarfs exceptionally challenging.  Low-mass companions with higher mass primaries for which ages can be more reliably determined are able to break this degeneracy, and are thus invaluable benchmarks for testing and empirically guiding substellar models (e.g., \citealt{faherty2010}).  

The coldest spectral class of brown dwarfs, the Y dwarfs \citep{cushing2011, kirkpatrick2012}, have estimated temperatures ($T_{\rm eff}$ $<$ 450 K) and masses ($\lesssim$20 M$_{\rm Jup}$) approaching those of the gas giants of our own Solar System.  The population of known Y dwarfs is scarce -- to date, only 25 Y-type objects have been spectroscopically confirmed \citep{cushing2011, kirkpatrick2012, liu2012, tinney2012, kirkpatrick2013, cushing2014, luhman2014a, pinfield2014, dupuy2015, schneider2015, leggett2017, martin2018, tinney2018}.  Benchmark companions at these extremely cold temperatures and low masses are rarer still.  Only one likely Y dwarf with a stellar-mass companion is known, WD 0806-661B, a companion to a DQ white dwarf at $\sim$19 pc \citep{luhman2011}.  As such, it is the only Y-type brown dwarf with well-constrained age and mass estimates (1.5$^{+0.5}_{-0.3}$ Gyr; \citealt{luhman2012}, 7.8$^{+1.0}_{-1.2}$ $M_{\rm Jup}$; \citealt{zhang2021}).  However, this object's extremely cold temperature (328$\pm$4 K; \citealt{zhang2021}) makes it so faint ($J$$\sim$25 mag; \citealt{luhman2014}) that obtaining its near-infrared spectrum has so far proven infeasible.  

The citizen science project Backyard Worlds: Planet 9\footnote{\url{https://backyardworlds.org}} (BYW; \citealt{kuchner2017}) has identified a new substellar companion that may bridge the gap between warmer, T-type companions (e.g., the T8.5 brown dwarf Wolf 940B; \citealt{burningham2009}) and WD 0806-661B. This discovery is Ross 19B, a companion to the nearby M dwarf Ross 19 that straddles the T/Y boundary.  In this paper, we describe the discovery of Ross 19B, as well as observations and analyses we have performed to characterize both the A and B components of this system.  

\section{Discovery of Ross 19B}
The BYW project engages volunteers to examine sets of {\it Wide-field Infrared Survey Explorer} ({\it WISE}; \citealt{wright2010}) images called ``flipbooks'' for candidate moving objects \citep{kuchner2017}.  With little formal training or background knowledge, citizen scientists can search for nearby, cold brown dwarfs with large proper motions.  To date, the BYW project has made numerous unique brown dwarf discoveries, including widely-separated companions and extremely cold brown dwarfs \citep{kuchner2017, debes2019, faherty2020, bardalez2020, schneider2020, meisner2020, kirkpatrick2020, peter2021, rothermich2021, meisner2021}. 

\begin{figure*}
\plotone{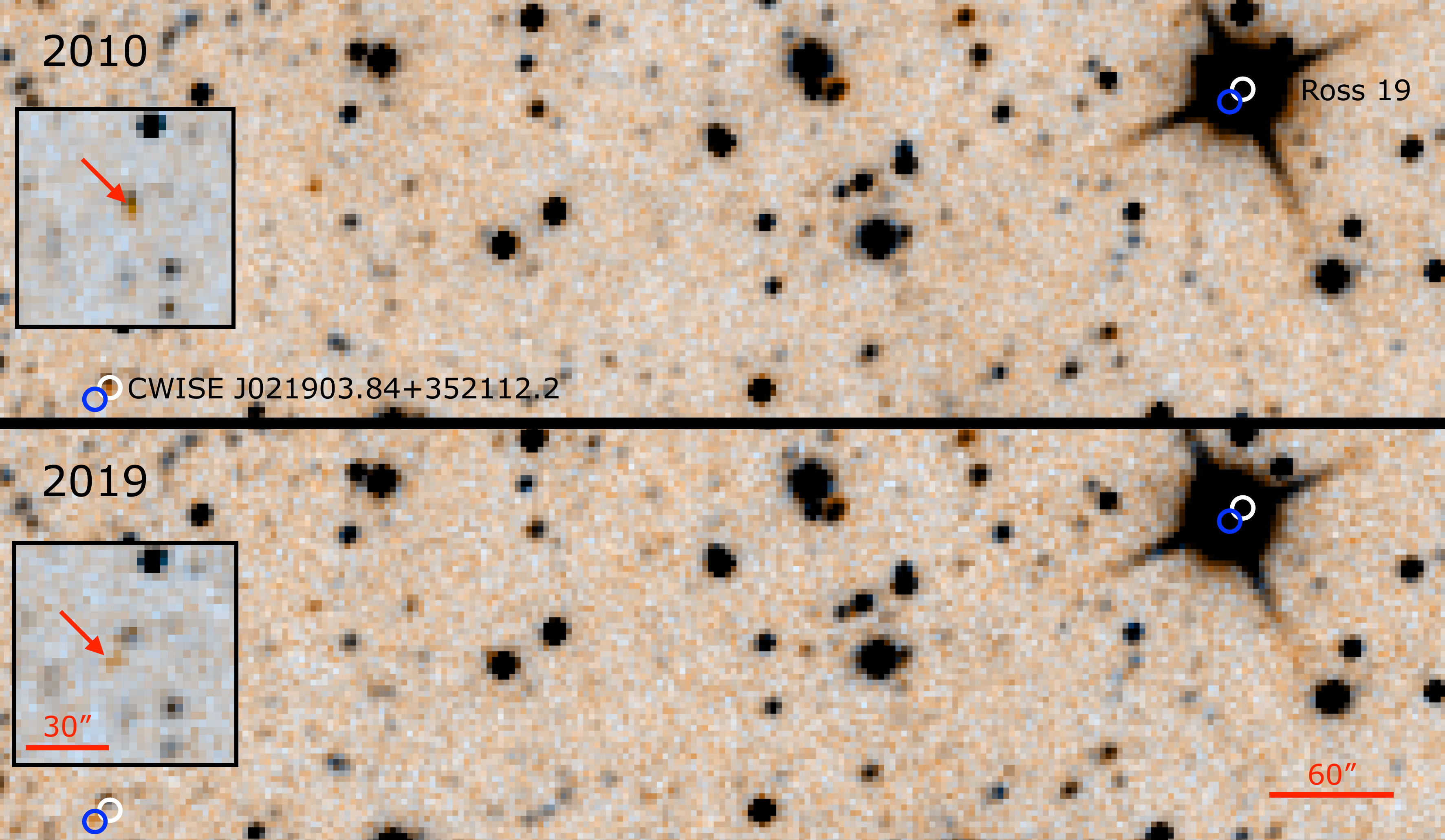}
\caption{unWISE images \citep{lang2014, meisner2018, meisner2019} showing the similar proper motions of Ross 19A (upper right) and CWISE J021948.68$+$351845.3 (lower left) from 2010 (upper panel) to 2019 (lower panel).  In these composite images, the W1 frames are color coded blue and the W2 frames are color coded red, and (W1+W2)/2 is coded green.  Note the orange color of CWISE J021948.68$+$351845.3, indicating that this object is significantly brighter at W2 (4.6 $\mu$m) than W1 (3.4 $\mu$m), a trait common to late-type brown dwarfs. The white circles indicate the 2010 positions of both objects, while the dark blue circles indicate their 2019 positions.  The inset images highlight the positions of CWISE J021948.68$+$351845.3. North is up and east is left for all images.}  
\label{fig:unWISE}
\end{figure*}

A fast-moving, cold brown dwarf candidate (CWISE J021948.68$+$351845.3) was identified through the BYW project by citizen scientists Samuel Goodman, L{\'e}opold Gramaize, Austin Rothermich, and Hunter Brooks at a large angular separation ($\sim$568\arcsec) from the nearby, high proper motion M dwarf Ross 19A (Figure \ref{fig:unWISE}).  The proper motion components of this source from the CatWISE 2020 catalog \citep{marocco2021} were noted to be similar to those of Ross 19A from the {\it Gaia} EDR3 catalog \citep{gaia2021}.  While similar, the $\mu_{\alpha}$ component of CWISE J021948.68$+$351845.3 differed from the {\it Gaia} EDR3 $\mu_{\alpha}$ measurement of Ross 19A by $\sim$1.5$\sigma$.  Further, the W1--W2 color\footnote{We use the CatWISE 2020 photometry that includes the proper motion solution ({\it w1mpro\_pm} and {\it w2mpro\_pm}), which is generally more accurate for moving sources than the stationary solution \citep{marocco2021}.} of CWISE J021948.68$+$351845.3 suggested a spectral type of $\sim$T8, which would indicate a distance several parsecs beyond that of Ross 19A.  However, we noted that in the earliest {\it WISE} epochs, CWISE J021948.68$+$351845.3 was partially blended with an unassociated background source (Figure 1), which could account for these astrometric and photometric discrepancies.  This source was not detected in the {\it WISE} All-Sky \citep{cutri2012} or AllWISE \citep{cutri2014} catalogs, most likely due to this blending at early {\it WISE} epochs.  We thus considered CWISE J021948.68$+$351845.3 as a potential companion to Ross 19A in need of additional astrometric confirmation.

\begin{deluxetable*}{lcccc}
\label{tab:ross19AB}
\tablecaption{Properties of Ross 19AB System}
\tablehead{
\colhead{Parameter} & \colhead{Value} & \colhead{Value} & \colhead{Ref.}}
\startdata
\cutinhead{Identifiers}
\dots & Ross 19A & Ross 19B & 1,4\\
CWISE & J021903.84$+$352112.2 & J021948.68$+$351845.3  & 2\\
\cutinhead{Observed Properties}
$\mu$$_{\alpha}$ (mas yr$^{-1}$) & 670.532 $\pm$ 0.042 & 673.2 $\pm$ 46.4 & 3,4\\
$\mu$$_{\delta}$ (mas yr$^{-1}$)  & -427.412$\pm$0.040 & -504.4$\pm$57.0 & 3,4\\
$\varpi$ (mas) & 57.3276$\pm$0.0398 & \dots & 3\\
$d$ (pc) & 17.444$\pm$0.012 & 17.58$\pm$3.75\tablenotemark{a} & 3, 4\\
RV (km s$^{-1}$) & -27.80$\pm$0.14 & \dots & 9\\ 
NUV (mag) & 22.281$\pm$0.249 & \dots & 5 \\
$G_{\rm Bp}$ (mag) & 12.822$\pm$0.003 & \dots & 3\\
$G$ (mag) & 11.388$\pm$0.003 & \dots & 3\\
$G_{\rm Rp}$ (mag) & 10.210$\pm$0.004 & \dots & 3\\
$J$ (mag) & 8.662$\pm$0.027 & 21.14$\pm$0.019 & 6,4\\
$H$ (mag) & 8.137$\pm$0.033 & \dots & 6\\
$K_{\rm S}$ (mag) & 7.897$\pm$0.024 & \dots & 6\\
W1 (mag) & 7.780$\pm$0.012 & 18.615$\pm$0.202 & 2\\
W2 (mag) & 7.545$\pm$0.008 & 15.810$\pm$0.055 & 2\\
W3 (mag) & 7.446$\pm$0.019 & \dots & 7\\
W4 (mag) & 7.169$\pm$0.096 & \dots & 7\\
\cutinhead{Inferred Properties}
Sp. Type &  M3.5 & T9.5$\pm$1.5 & 8,4\\
$T_{\rm eff}$ (K)  & 3481$\pm$49 & 500$^{+115}_{-100}$ &  4\\
log($L_{\rm bol}$/$L_{\odot}$) & -1.799$\pm$0.093 & \dots & 4\\
$R$ ($R_{\rm Jup}$) & 3.38$\pm$0.03 & \dots & 4\\
$M$ ($M_{\odot}$) & 0.362$\pm$0.007 & 0.015--0.038 & 4\\
{[}Fe/H] (dex) & -0.40$\pm$0.12 & \dots & 4\\
Age (Gyr) & 7.2$^{+3.8}_{-3.6}$ & \dots & 4 \\
\enddata
\tablenotetext{a}{Estimated using relations in \cite{kirkpatrick2020}.}
\tablerefs{ (1) \cite{ross1925}; (2) CatWISE 2020 \citep{marocco2021}, (3) {\it Gaia} EDR3 \citep{gaia2021}, (4) this work; (5) \cite{jones2016}; (6) 2MASS \citep{skrutskie2006}; (7) AllWISE \citep{kirkpatrick2014}; (8) \cite{bidelman1985}; (9) \cite{jeffers2018}}
\end{deluxetable*}

\section{Observations}

\subsection{Ross 19B}
\subsubsection{Keck/MOSFIRE}

\begin{figure}
\plotone{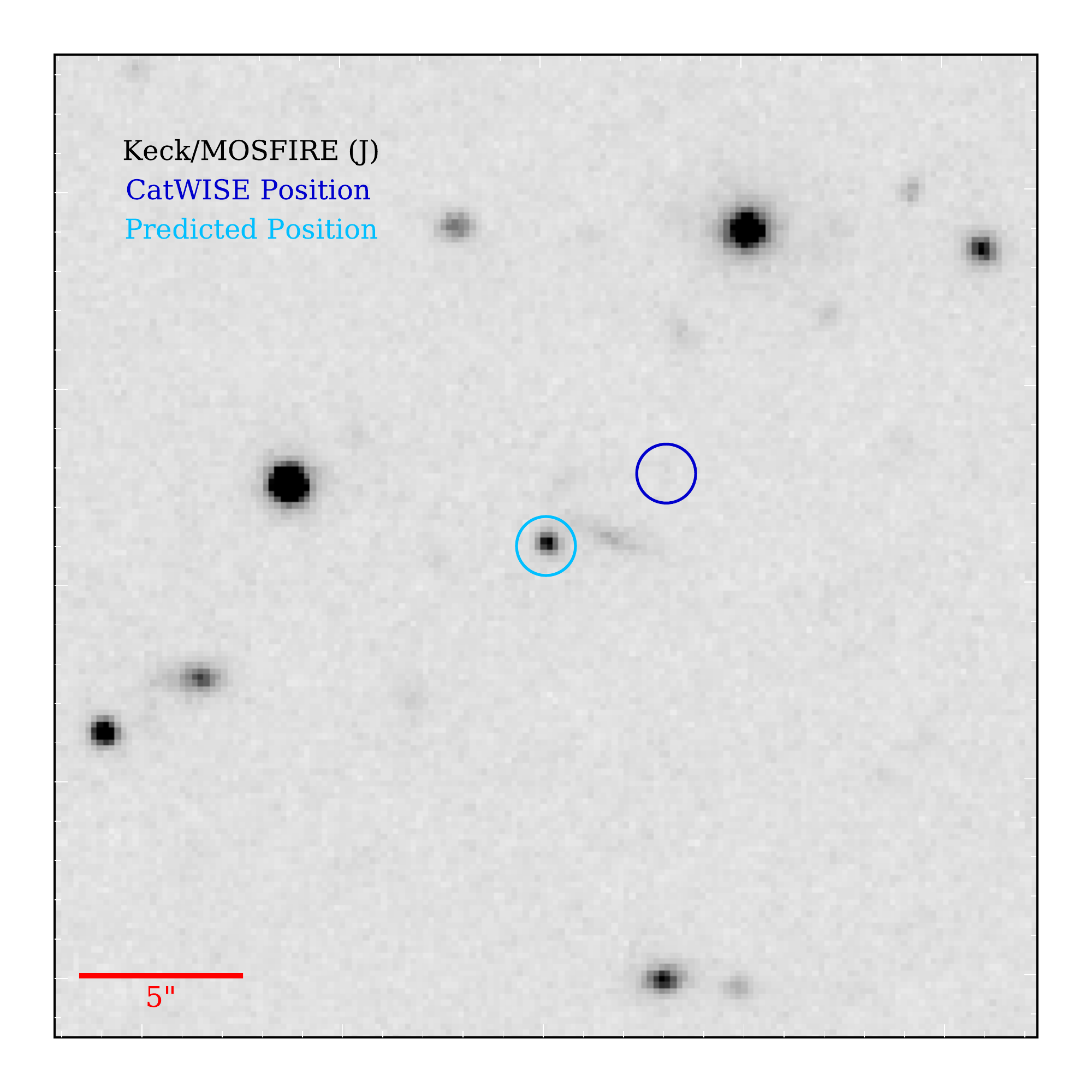}
\caption{Keck/MOSFIRE image of CWISE J021903.84$+$352112.2, showing its original CatWISE 2020 position (dark blue circle), and its predicted position based on the {\it Gaia} EDR3 proper motions of Ross 19A (light blue).}  
\label{fig:mosfire}
\end{figure}

To assess the possible association of CWISE J021948.68$+$351845.3 and Ross 19A, we obtained a deep, $J-$band follow-up image using the Multi-Object Spectrometer For Infra-Red Exploration (MOSFIRE; \citealt{mclean2012}) with Keck on UT 2020 September 4 (Figure \ref{fig:mosfire}).  MOSFIRE has a 6.1\arcmin\ $\times$ 6.1\arcmin\ field of view with a pixel scale of 0\farcs1798 pixel$^{-1}$.  Observing conditions were ideal, with a clear sky and a seeing of $\sim$0\farcs5. The $J-$band stacked image was obtained by acquiring 18$\times$100s frames on a 9-point dithering pattern. The images were reduced with custom IDL routines, using bright, unsaturated stars for photometric calibration, whose 2MASS $J$ magnitudes were converted to the MKO system with the equations provided at NASA's Infrared Science Archive\footnote{\url{https://irsa.ipac.caltech.edu/data/2MASS/docs/releases/allsky/doc/sec6\_4b.html}}. Magnitudes were measured using a 3\arcsec\ circular aperture with a sky annulus of inner radius 3\farcs2 and outer radius 5\farcs7. The aperture correction was estimated using the same set of bright, unsaturated stars used for photometric calibration.  We measured a $J-$band magnitude of 21.14$\pm$0.02 mag for CWISE J021948.68$+$351845.3. These data allowed us to both refine this object's astrometry and its spectral type estimate (see Section \ref{sec:astrometry} and Section \ref{sec:B_spt}).  As seen in Figure \ref{fig:mosfire}, the predicted position of this object using the {\it Gaia} EDR3 astrometry from Ross 19A is fully consistent with its position at the time of the MOSFIRE image, providing further evidence that these objects are physically associated.  

\subsection{Ross 19A}

With the knowledge that CWISE J021948.68$+$351845.3 was a possible companion to Ross 19A, we also sought to characterize the stellar component in this potential system with optical and near-infrared spectroscopy.

\subsubsection{Lick/KAST}
A red optical spectrum of Ross 19A was obtained with the KAST spectrograph on the Lick 3m Shane Telescope on 14 December 2020 (UT). Observations were obtained through light cirrus clouds with variable seeing of 2$\arcsec$-2$\farcs$5. Two exposures of 250~s were obtained with the 2$\arcsec$ slit and 600/7500 red grating, providing resolution $\lambda/\Delta\lambda$ $\approx$ 1800 spanning 6300$-$9000~{\AA}. The G2~V star HD~12846 ($V$ = 6.89 mag) was observed immediately afterward for telluric calibration, and the flux standard Hiltner~600 was observed during the night for flux calibration \citep{hamuy1994}. We also obtained flat field and HeHgNe arc lamps exposures at the start of the night for pixel response and wavelength calibration. Data were reduced using the \texttt{kastredux} package\footnote{\url{https://github.com/aburgasser/kastredux}} using default settings.

\begin{figure*}
\plotone{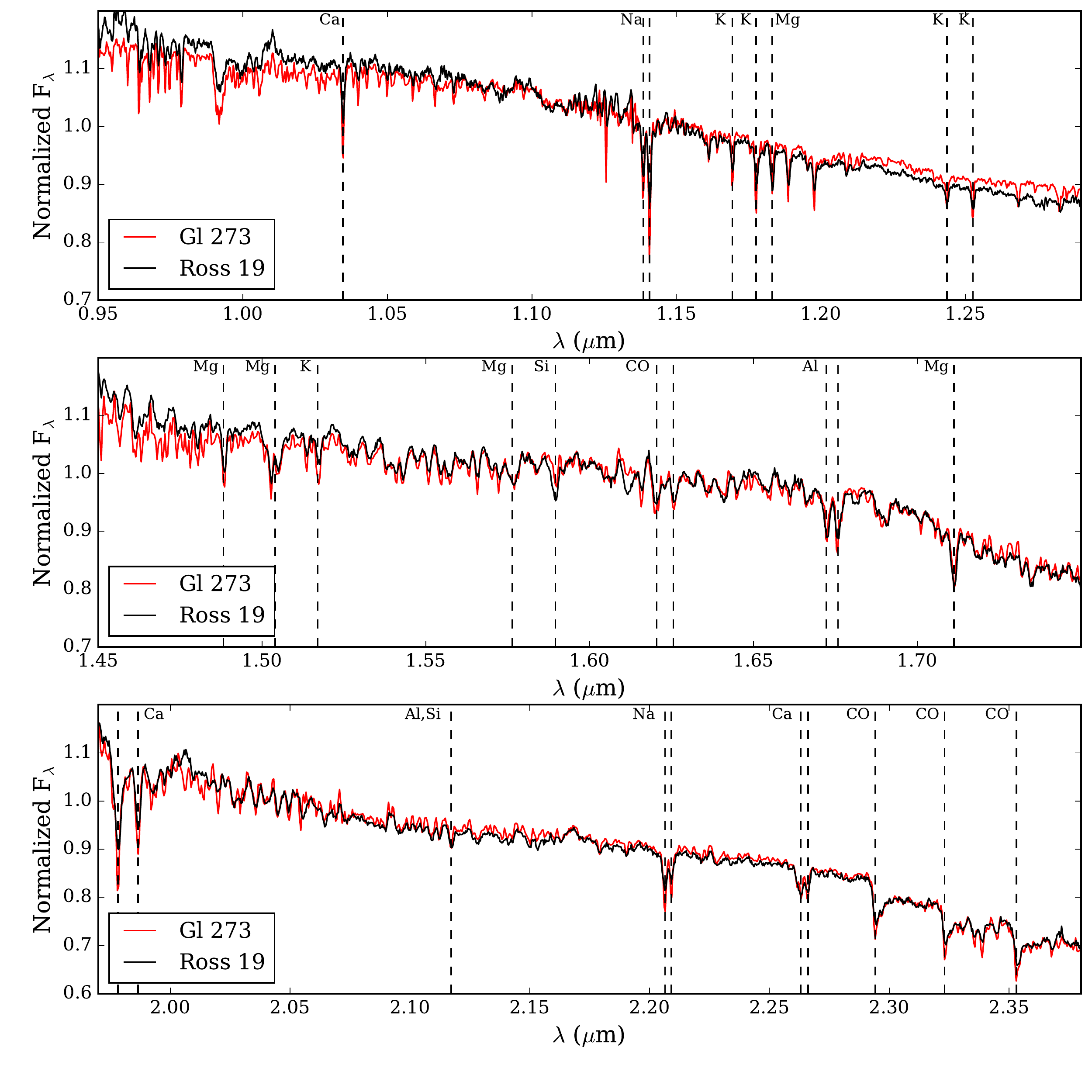}
\caption{IRTF/SXD $J-$band (top) $H-$band (middle), and $K-$band (bottom) spectrum of Ross 19A compared to Gl 273 (Luyten's Star), which has a similar spectral type.  Prominent absorption features are labeled.}  
\label{fig:spex}
\end{figure*}

\subsubsection{IRTF/SpeX}
We obtained a near-infrared spectrum of Ross 19A with the SpeX spectrograph \citep{rayner2003} at NASA's 3 m Infrared Telescope Facility (IRTF) on UT 2020 Dec 23.  The observations were taken in short cross-dispersed (SXD) mode, which gives a resolution of $\lambda/\Delta\lambda$ $\approx$750 across the 0.8$-$2.4 $\mu$m wavelength range. The A0 star HD 13869 was observed immediately after Ross 19A for telluric correction purposes.  Conditions were clear and Ross 19A was the first target of the night when the seeing was measured at 0\farcs45 at $K$.  We took 4 images of 90 seconds each in an ABBA pattern for the target with the 0\farcs8 slit and 12 images of 1 second for the telluric standard.  For both observations, the slit was aligned to the parallactic angle.  Calibration files (flats and arcs) were taken between the target and the telluric observations. Spectral extraction and telluric correction were performed with the SpeXTool package \citep{vacca2003, cushing2004}. The reduced spectrum is shown in Figure \ref{fig:spex} compared to the similar spectral type object Gl 273 from the IRTF spectral library \citep{cushing2005, rayner2009}.  As seen in the figure, Ross 19's spectrum is rich in atomic lines, many of which are excellent diagnostics of M-type star physical properties (see Section \ref{sec:ross19a}). 

\section{Analysis}

\subsection{Astrometry of CWISE J021903.84$+$352112.2}
\label{sec:astrometry}

Because CWISE J021903.84$+$352112.2 is blended with a background object in some of the first {\it WISE} epochs, we independently measured the proper motion of this source. To do this, we first re-registered the world coordinate system (WCS) of the Keck/MOSFIRE image with sources from the {\it Gaia} EDR3 catalog \citep{gaia2021}.  We similarly used {\it Gaia} EDR3 to re-register unWISE yearly W2 coadds \citep{lang2014, meisner2018, meisner2019}, with seven 1-year coadds from 2010--2011 to 2018--2019. We ran the \texttt{crowdsource} source extraction code \citep{schlafly2018, schlafly2019}, designed to perform point-source photometry in crowded regions, on each unWISE epochal coadd.  Because \texttt{crowdsource} has been shown to work well for blended or partially blended sources in {\it WISE} images \citep{schlafly2018, schlafly2019}, it is well-suited for this object.  We measured proper motion components of $\mu_{\alpha}$ = 673.2$\pm$46.4 and $\mu_{\delta}$ = $-$504.4$\pm$57.0 mas yr$^{-1}$ using unWISE and Keck/MOSFIRE positional measurements for CWISE J021903.84$+$352112.2.  A comparison of these values to those of Ross 19A from the {\it Gaia} EDR3 catalog gives differences of 0.06$\sigma$ and 1.35$\sigma$, respectively.  Astrometry for both sources is provided in Table \ref{tab:ross19AB}.  

\subsection{Chance Alignment Probability}

The similarity of the measured proper motion components for CWISE J021903.84$+$352112.2 and Ross 19A are strongly suggestive of a co-moving, physically associated pair.  Further, using the \cite{kirkpatrick2020} W2 to {\it Spitzer} ch2 relation and the absolute $J$ versus $J-$ch2 color relation for known T and Y dwarfs with measured parallaxes from \cite{kirkpatrick2020}, we estimate a distance to CWISE J021903.84$+$352112.2 of 17.58$\pm$3.75 pc, fully compatible with the {\it Gaia} EDR3 distance of Ross 19A (17.444 $\pm$ 0.012 pc; \citealt{gaia2021}).  

We estimated a co-moving probability between the host star and companion with \texttt{CoMover} \citep{gagne2021}, a custom-written IDL wrapper code that uses the engine of the BANYAN~$\Sigma$ software used to determine the probability that a given star is a member of a nearby young association \citep{gagne2018}, using Bayesian statistics. BANYAN~$\Sigma$ uses the sky position, proper motion, and optionally the heliocentric radial velocity and parallax of a star to determine how well it matches the Galactic coordinates ($XYZ$) and space velocities ($UVW$) of each young association and unrelated stars in the neighborhood of the Sun. The model that represents each of these hypotheses consists of a multivariate Gaussian model in 6-dimensional space; a single Gaussian component is used for each association, and a 10-component model is used to represent the non-Gaussian $XYZUVW$ distribution of nearby field stars. BANYAN~$\Sigma$ has two specific advantages that can also benefit the calculation of co-moving probabilities: (1) the code can calculate a probability even if the star being tested does not have heliocentric radial velocity or parallax measurements, by marginalizing Bayes' theorem over these dimensions; and (2) the marginalization integrals were solved analytically for multivariate Gaussian models, making the code much faster and robust against numerical inaccuracies. \cite{gagne2018} provide detailed information on the core functioning of the BANYAN~$\Sigma$ engine, and the code is publicly available on GitHub\footnote{\url{https://github.com/jgagneastro/CoMover}}.

The problem of estimating a co-moving probability bears some analogy with the goals of BANYAN~$\Sigma$. If the full $XYZUVW$ position of the host star is known, it can be modelled with a single 6-dimensional multivariate Gaussian, and observables of the potential companion can be tested against this model and the BANYAN~$\Sigma$ model of field stars with the BANYAN~$\Sigma$ engine directly. In cases where the full kinematics of the host star are not known, a series of discrete multivariate Gaussian models can be used in the BANYAN~$\Sigma$ engine, and then numerically marginalized by summing over the membership probabilities across all host-star models. We have currently only implemented such numerical marginalization over unknown host-star heliocentric radial velocities in the \texttt{CoMover} wrapper.

There are two further complications in the determination of co-moving probabilities. First, a physical size must be chosen for the $XYZ$ spatial part of the host-star model. Here, we decided to use a model with a characteristic width of 0.1 pc, corresponding to physical separations of about 20\,600 au, larger than some of the widest-separation binaries known to be at least lightly gravitationally bound \citep{caballero2012, deacon2014, marocco2020}. This choice for the spatial size of the model should therefore be applicable to systems out to the extreme edge of known bound companions.  Some seemingly co-moving and coeval stars with separations beyond 0.1 pc have also been reported (e.g., see \citealp{dhital2010, oh2017, nelson2021}), however they likely to correspond to members of whole or partially dissolved coeval young associations, rather than being gravitationally bound. Therefore, by choosing a spatial model with a size of 0.1 pc, we are specifically ignoring these potentially coeval but non-gravitationally bound populations of stars.

The second complication relates to small but statistically significant differences in $UVW$ space velocities that are due to orbital motion. Such differences have recently been used to measure the masses of directly-imaged exoplanets given the exquisite precision of Gaia data (e.g., see \citealt{brandt2021}), and were also shown to cause discrepancies in Gaia-based $UVW$ velocities of an M dwarf/white dwarf pair \citep{peter2021}. We use the projected physical separation between a host star and its companion to estimate the orbital velocity of a circular orbit to account for this effect in a way that still allows using the analytical solutions of BANYAN~$\Sigma$. This requires using a rough estimate of the host-star mass, which we combine with the projected physical separation of the companion and Kepler's Third Law to determine a worst-case-scenario difference in $UVW$ caused by such orbital motion. The resulting orbital motion was chosen as the characteristic width of the $UVW$ kinematic part of the host-star model, to which we added in quadrature a slightly conservative 1 km s$^{-1}$ to account for combined effects of gravitational reddening and convective blueshift (e.g., \citealt{dai2018, lohner2019}) that put a lower limit on the measurement error of the heliocentric velocity of the host star.

We have left the Bayesian priors of our framework to unity for the sum over all host-star models, and to the local field density \citep{kirkpatrick2012} for the field-stars model. It would be possible to also include considerations of binary fraction in these Bayesian priors, but it would be best to do so in a way that depends on the host star's spectral type. Future improvements could also include a non-Gaussian spatial shape for the host-star model that captures the distribution of known companions as a function of physical separation.

\begin{figure*}
\plotone{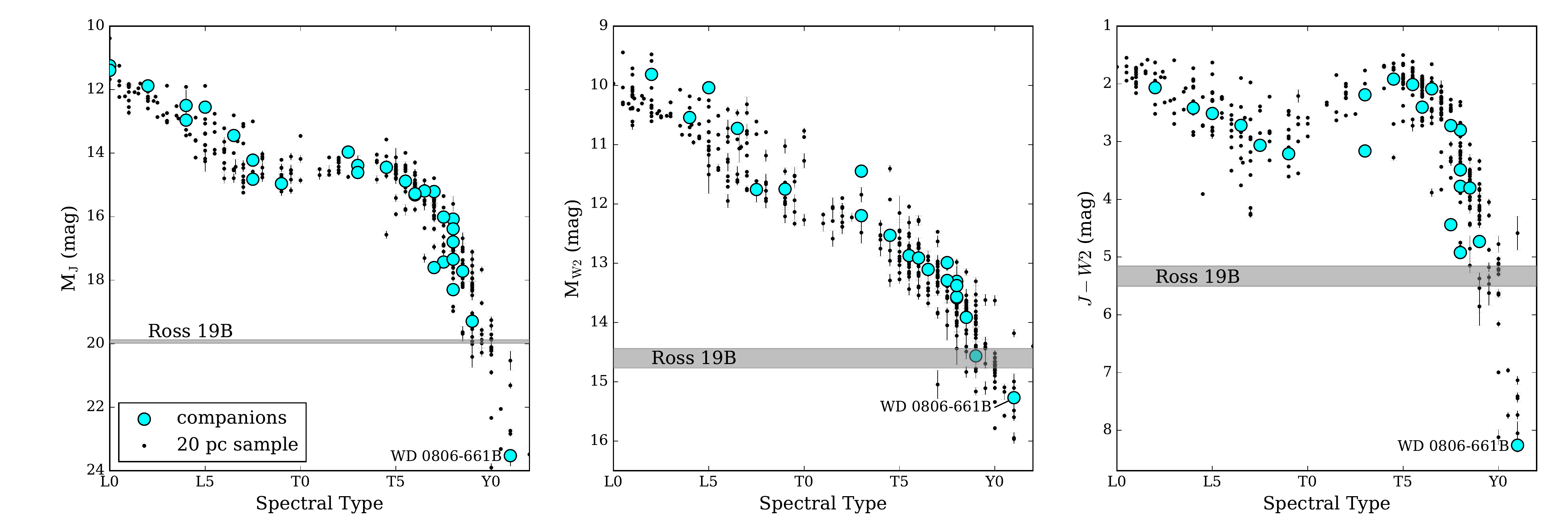}
\caption{Absolute $J$ magnitude, absolute W2 magnitude, and $J-$W2 color versus spectral type for the 20 pc sample of nearby L, T, and Y dwarfs from \cite{kirkpatrick2020}.  Known wide companions are highlighted in light blue.  The 3$\sigma$ absolute magnitudes (based on the {\it Gaia} EDR3 distance to Ross 19A) and color of Ross 19B are indicated by the gray shaded region in each panel. The position of the cold companion WD 0806-661B in these diagrams (labeled) is discussed in the text. } 
\label{fig:cmds}
\end{figure*}

Using our updated astrometry for CWISE J021903.84$+$352112.2, the {\it Gaia} EDR3 astrometry of Ross 19A combined with the radial velocity from \cite{jeffers2018} (-27.8$\pm$0.14 km s$^{-1}$), and the \texttt{CoMover} wrapper, we find a 100\% probability that these objects are physically associated.  

As an additional test to validate the result of our novel CoMover routine, we slightly perturb CWISE J021903.84$+$352112.2 in $UVW$ space using its proper motion and distance uncertainties, then compare the best-case $UVW$ separation between Ross 19 and the field star models.  We repeated this process 1000 times in a Monte Carlo fashion, and find a maximum-likelihood $UVW$ separation of 7$^{+4}_{-3}$ km s$^{-1}$ for Ross 19, compared to 40$\pm$4 km s$^{-1}$ for the field. This supports the CoMover result that Ross 19 is favored over the field star model.  These uncertainties will be reduced as the astrometry of CWISE J021903.84$+$352112.2 is further refined.    

While our novel method of testing co-moving companionship provides strong evidence of an associated pair, we also ensured that these objects satisfy wide binary conditions defined in previous studies.  Ross 19A and CWISE J021903.84$+$352112.2 satisfy the criteria for evaluating common proper motion pairs found in \cite{lepine2007b}, \cite{dupuy2012}, and \cite{deacon2014}.  We therefore designate CWISE J021903.84$+$352112.2 as Ross 19B.

\subsection{Spectral Type of Ross 19B}
\label{sec:B_spt}

While a definitive spectral type will require future spectroscopic observations from a facility such as {\it HST} or {\it JWST}, we can provide an estimate of Ross 19B's spectral type based on its available photometry and the distance to Ross 19A.  The following analysis uses the empirical relations from \cite{kirkpatrick2020}.  To compare Ross 19B to other cold brown dwarfs with measured distances, we first make the small conversion from {\it WISE} W2 magnitude to {\it Spitzer Space Telescope} ch2 magnitude, and find ch2 = 15.765$\pm$0.115 mag.  Then, using the $J-$ch2 color versus spectral type relation, we find a photometric spectral type estimate of 19.6 (where T0 = 10 and Y0 = 20), with a systematic uncertainty of 0.53 subtypes.  

As a second spectral type estimate, we use the {\it Gaia} EDR3 parallax of Ross 19A, the $J-$band and W2 magnitudes of Ross 19B with the absolute magnitude versus spectral type relations of \cite{kirkpatrick2020}. We find photometric spectral type estimates of 19.6$\pm$0.6 and 19.9$\pm$1.3 for $M_{\rm J}$ and $M_{\rm W2}$, respectively. All available information points to a spectral type near the T/Y boundary for Ross 19B, and we assign a conservative spectral type estimate of T9.5$\pm$1.5.  This spectral type range corresponds to an effective temperature (\teff) of 500$^{+115}_{-100}$ K \citep{kirkpatrick2020}.

In Figure \ref{fig:cmds}, we show $M_{\rm J}$, $M_{\rm W2}$, and $J-$W2 versus spectral type for the 20 pc sample of L, T, and Y dwarfs from \cite{kirkpatrick2020} compared to Ross 19B.  We also include the recently discovered T9 companion COCONUTS-2b \citep{zhang2021b}. This figure highlights that, no matter the spectral type of Ross 19B, it is likely colder than all known wide companions within 20 pc with the exception of WD 0806-661B. WD 0806-661B does not have a measured spectral type or a $J-$band magnitude.  To include this object in diagrams in Figure \ref{fig:cmds}, we converted its {\it HST} F110W magnitude from \cite{luhman2014} to an approximate $J$ magnitude by finding the synthetic $J-$F110W colors of known Y dwarfs with spectra from \cite{schneider2015} that cover the entire F110W filter. Known Y dwarfs are $\sim$0.75 mag brighter in $J$ than F110W, giving WD 0806-661B a $J$-mag of $\sim$25.0.  We also adopt the photometric spectral type estimate of Y1 from \cite{kirkpatrick2020} for WD 0806-661B.  

\begin{figure*}
\plotone{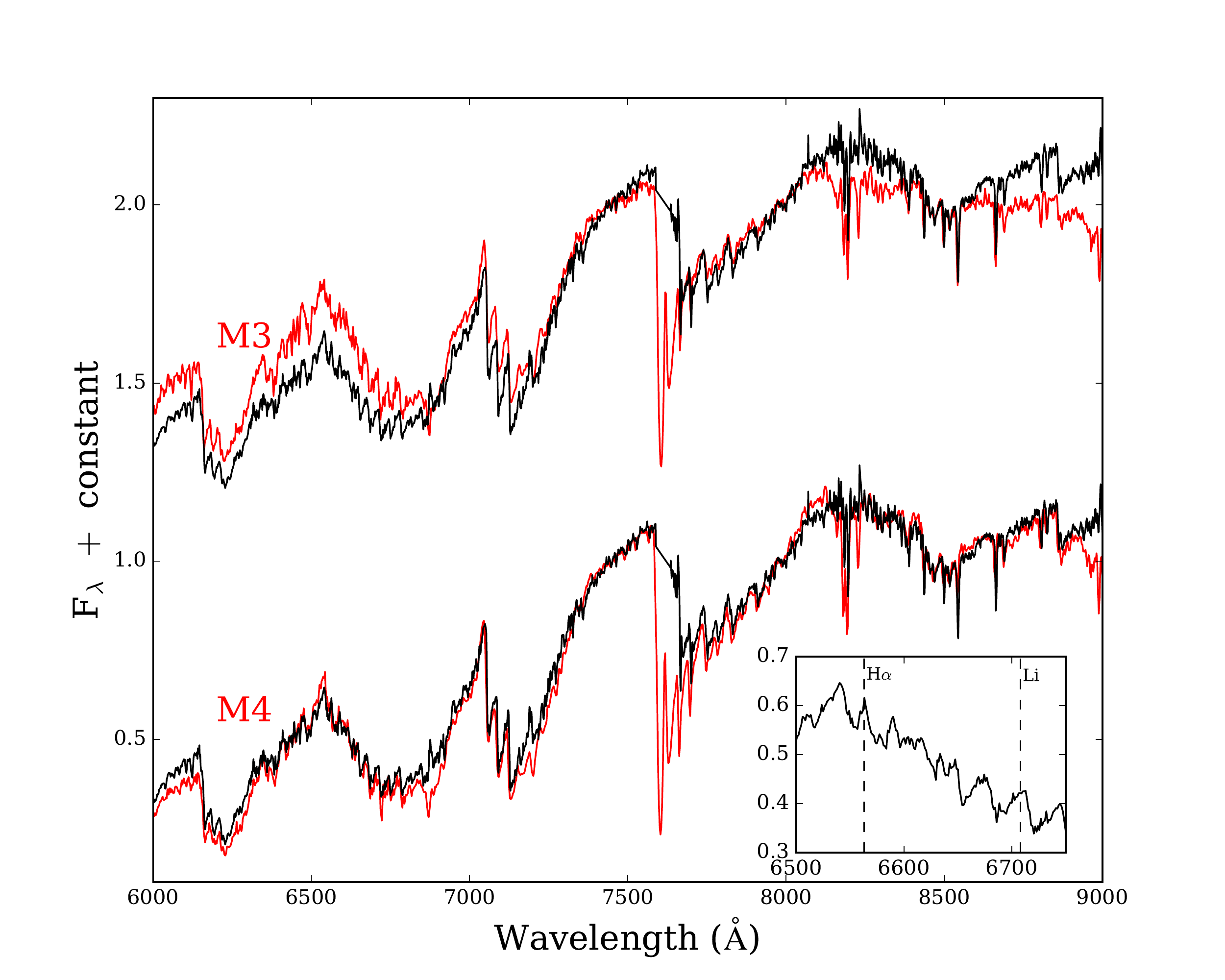}
\caption{Lick/KAST spectrum of Ross 19A compared to the M3 and M4 spectral standards from \cite{kirkpatrick1991}.  The inset highlights the weak H$\alpha$ emission of Ross 19A and the lack of lithium absorption at $\lambda$6707$\AA$.} 
\label{fig:kast}
\end{figure*}

\subsection{Properties of Ross 19A}
\label{sec:ross19a}

\subsubsection{Spectral Type}
Ross 19A was discovered by \cite{ross1925} and was spectroscopically classified as M3.5 by \cite{bidelman1985}.  We confirm the M3.5 spectral type designation using our Lick/KAST optical spectrum by comparing to M dwarf optical standards \citep{kirkpatrick1991}, shown in Figure \ref{fig:kast}.

\subsubsection{Basic Properties: Effective Temperature, Luminosity, Radius, Mass}

To estimate basic physical properties of Ross 19A, we follow the method of \cite{filip2015} and use the SEDkit package \citep{filip2020} whereby photometry, spectroscopy, and a parallax are combined to calculate a bolometric luminosity.  Combining this luminosity with an age estimate, we use the Dartmouth Magnetic Evolutionary Stellar Tracks and Relations (DMEstar) models \citep{feiden2012, feiden2013} to obtain a radius estimate.  We then apply Wien's Law to semi-empirically measure the effective temperature.

\begin{figure*}
\plotone{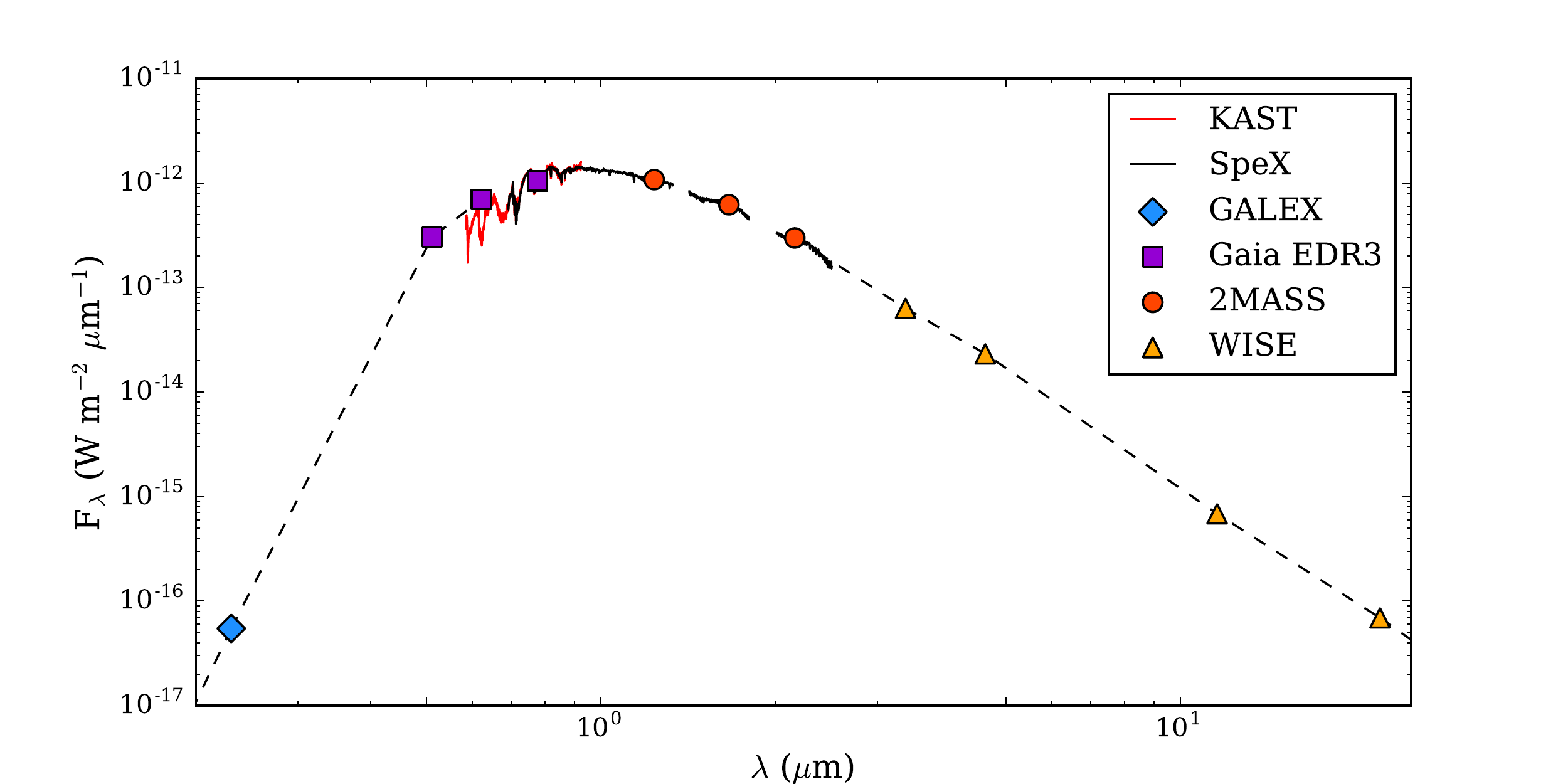}
\caption{Spectral energy distribution for Ross 19A.  Our KAST and SpeX spectra for Ross 19A are plotted as red and black lines, respectively.  All photometry is labeled as in the figure legend and includes GALEX NUV, {\it Gaia} $G_{\rm Bp}$, $G$, $G_{\rm Rp}$, 2MASS $J$, $H$, $K_{\rm S}$, and {\it WISE} W1, W2, W3, and W4 magnitudes, . This comprehensive SED is used to determine fundamental parameters of Ross 19A following the methods in \cite{filip2015}.} 
\label{fig:sed}
\end{figure*}

To construct a comprehensive spectral energy distribution (SED) for Ross 19A, we use our optical KAST spectrum, our near-infrared SpeX spectrum, and photometry from various sky surveys listed in Table \ref{tab:ross19AB}. Note that while Ross 19A is within the Pan-STARRS survey footprint \citep{kaiser2010}, it is brighter than the Pan-STARRS saturation limit and therefore not included in our SED. The full SED for Ross 19A is shown in Figure \ref{fig:sed}.  The values for the luminosity, \teff, mass, and radius of Ross 19A we find using the \cite{filip2015} approach are listed in Table \ref{tab:ross19AB}.    

As a sanity check on the results from our SED analysis of Ross 19A, we also estimate the properties of this star using the \cite{newton2015} empirical relations.  These relations connect luminosity and temperature with $H-$band Al and Mg features.  The \cite{newton2015} relations (their equations 1 and 3) give $T_{\rm eff}$ = 3409$\pm$73 K and log $L/L_{\odot}$ = -1.90$\pm$0.05, fully consistent with the values from our SED analysis, which we take as our final values.  These $T_{\rm eff}$ estimates also compare well with those from previous studies: 3280 K \citep{lepine2013}, 3433$\pm$ 42 K \citep{gaidos2014},  3354$\pm$72 K \citep{muirhead2018}, 3366 K \citep{houdebine2019}, and 3384$\pm$104 K \citep{sebastian2021}.  

\subsubsection{Metallicity}
\label{sec:metal}

Historically, determining M dwarf metallicities has been challenging because of the abundance of complex molecular absorption bands and the lack of a recognizable continuum.  However, the strengths of some broad features, such as TiO and CaH bands, change significantly with metallicity.  This led to broad metallicity classes: normal dwarfs (dM), subdwarfs (sd), extreme subdwarfs (esd), and ultra subdwarfs (usd) \citep{gizis1997, lepine2007} based on spectroscopic indices of various molecular bands.  We measured TiO5 = 0.484, CaH2 =  0.496, and CaH3 = 0.766, which we find to be consistent with a normal M dwarf \citep{lepine2007,zhang2019}. This is consistent with \cite{lepine2013}, who found a metallicity index approximately solar for Ross 19A.

While refinements of metallicity classes for M dwarfs are ongoing \citep{lodieu2019, zhang2019}, there has also been a concerted effort to identify metal-sensitive features in the near-infrared spectra of M-type stars. \cite{covey2010} defined the H$_2$O-H and H$_2$O-K spectral indices to characterize near-infrared water absorption bands for young stellar objects (YSOs).  \cite{rojas2010} used the \cite{covey2010} H$_2$O-K index and $K-$band Na I and Ca I absorption features to create an empirical relation with [Fe/H] values for nearby M dwarfs.

\cite{mann2013a} used a large sample of M-type companions to F, G, and K-type host stars with known metallicities to show that the most metal-sensitive features for M-type stars occur in the $K-$band portion of the near-infrared spectrum.  For this reason, we use the \cite{mann2013a} $K-$band [Fe/H] relation for early M stars to determine the metallicity of Ross 19A.  Following \cite{rojas2012}, equivalent widths are approximated using

\begin{equation}
    {\rm EW}_{\lambda} \simeq \sum_{i=0}^{n-1} \left[ 1-\frac{F(\lambda_i)}{F_c(\lambda_i)} \right] \Delta\lambda_i,
\end{equation}

\noindent where $\lambda_i$ is the wavelength at pixel $i$,  $F(\lambda_i)$ is the flux at $\lambda_i$, and $F_c(\lambda_i)$ is the estimated continuum flux at $\lambda_i$.  Following \cite{mann2013a}, we interpolate the spectrum to a higher resolution ($\sim$10000) to account for finite pixel sizes.  For a consistency check, we measured [Fe/H] values for several M dwarfs in the IRTF spectral library \citep{cushing2005, rayner2009} that also have measurements in \cite{newton2014}, and we find our measured values to be consistent with an rms scatter of 0.12 dex. We take this as our measurement uncertainty.      

We calculate a metallicity below solar for Ross 19A ([Fe/H] = $-$0.40$\pm$0.12 dex). This makes Ross 19B one of very few low-metallicity benchmark cold companions, which include the T8 companion Wolf 1130C ([Fe/H] = $-$0.70$\pm$0.12; \citealt{mace2013, mace2018}) and the T8 companion BD+01\degr 2920B ([Fe/H] = $-$0.38$\pm$0.06; \citealt{pinfield2012}), both of which are likely warmer than Ross 19B by $\gtrsim$100 K ($\sim$650 K; \citealt{zhang2021}).  

The near-infrared metallicity determination for Ross 19A is in some tension with the metallicity inferred from its optical spectrum.  \cite{mann2013a} noted several similar cases in which M-type companions with low-metallicity primaries were classified as approximately solar metallicity based on their optical spectra.  $K-$band metallicities were found to be in much better agreement with the metallicities inferred from their primary stars.  We therefore take the metallicity measured using the $K-$band for Ross 19A as our final value.  Further investigations between optical and near-infrared metallicity determinations for low-metallicity M-type stars may help to resolve this tension.     

\subsubsection{Age}

M dwarfs are generally difficult to age-date because their lifetimes are so long that they do not evolve appreciably on the main sequence.  Therefore the common practice of isochrone-fitting based on color-magnitude diagram positions is usually not applicable. To constrain the age of the Ross 19AB system, we have investigated multiple diagnostics for Ross 19A, including color-magnitude diagram position, rotation, activity, metallicity, and kinematics. 

The position of Ross 19A on a color-magnitude diagram compared to the well-characterized spectroscopic M dwarf sample of \cite{kiman2019} shows it is a typical mid-type M dwarf, with no clear deviation above or below the main sequence.   

We measure an H$\alpha$ equivalent width of $-$0.62$\pm$0.03 \AA\ for Ross 19A from our Lick/KAST spectrum, consistent with the $-$0.57 \AA\ H$\alpha$ measurement in \cite{lepine2013}.  This H$\alpha$ emission is relatively weak for an object with an M3.5 spectral type \citep{newton2017, jeffers2018, kiman2021}. This indicates that Ross 19B has an age greater than $\approx$600--800 Myr, as all mid-M type members of the Hyades and Praesepe clusters show strong H$\alpha$ emission \citep{douglas2014, fang2018}.  We also note here that there is no notable lithium absorption at $\lambda$6707 \AA\ in the spectrum of Ross 19A.  This is not surprising, because Ross 19A would need to be exceptionally young to have detectable lithium, as even the majority of $\sim$M3.5 members of the $\sim$24 Myr old $\beta$ Pictoris Moving Group do not have detectable lithium (e.g., \citealt{messina2016, shkolnik2017}). Ross 19A also has a low-signal to noise ratio (S/N) near-UV (NUV) detection (22.281$\pm$0.249 mag) from the {\it Galaxy Evolution Explorer} ({\it GALEX}).  This value is consistent with a field-age star exhibiting low activity \citep{jones2016, schneider2018}.   

\cite{jeffers2018} ruled out Ross 19A as a fast rotator, finding a $v$sin$i$ upper limit of 3.0 km s$^{-1}$.  We searched available photometric archives for evidence of rotational modulations, and found no clear variability in the light curves from the Transiting Exoplanet Survey Satellite (TESS; \citealt{ricker2015}) or from the Zwicky Transient Facility (ZTF; \citealt{bellm2019}).  This is consistent with the results of \cite{see2021}, who showed a positive correlation between variability amplitude and metallicity (see Section \ref{sec:metal}).  We also see no evidence of flares in either light curve, again consistent with an age of several Gyr (e.g., \citealt{davenport2019}).

Even though correlations between age and metallicity are generally weak, we can use the subsolar metallicity of Ross 19A found in Section \ref{sec:metal} to place broad constraints on its age using catalogs of stars with known ages and metallicities following the formalism of \cite{burgasser2017}.  In that work, the metallicity of TRAPPIST-1 was compared to the metallicity- and age-calibrated samples from the Spectroscopic Properties of Cool Stars (SPOCS) survey \citep{valenti2005} and the  Geneva-Copenhagen Survey (GCS) \citep{casagrande2011}. In this work, we expand the stellar samples to include \cite{bensby2014},  \cite{brewer2016}, and \cite{luck2017, luck2018}, in addition to the SPOCS and GCS samples. Stellar ages in the \cite{valenti2005}, \cite{bensby2014}, and \cite{brewer2016} samples were found using the Yonsei-Yale (Y2) isochrones \citep{demarque2004}, while \cite{casagrande2011} uses Padova isochrones \citep{bertelli2008} and the studies of \cite{luck2017, luck2018} uses a combination of Y2, Dartmouth \citep{dotter2008} and BaSTI isochrones \citep{basti2016}.   

\begin{figure*}
\plotone{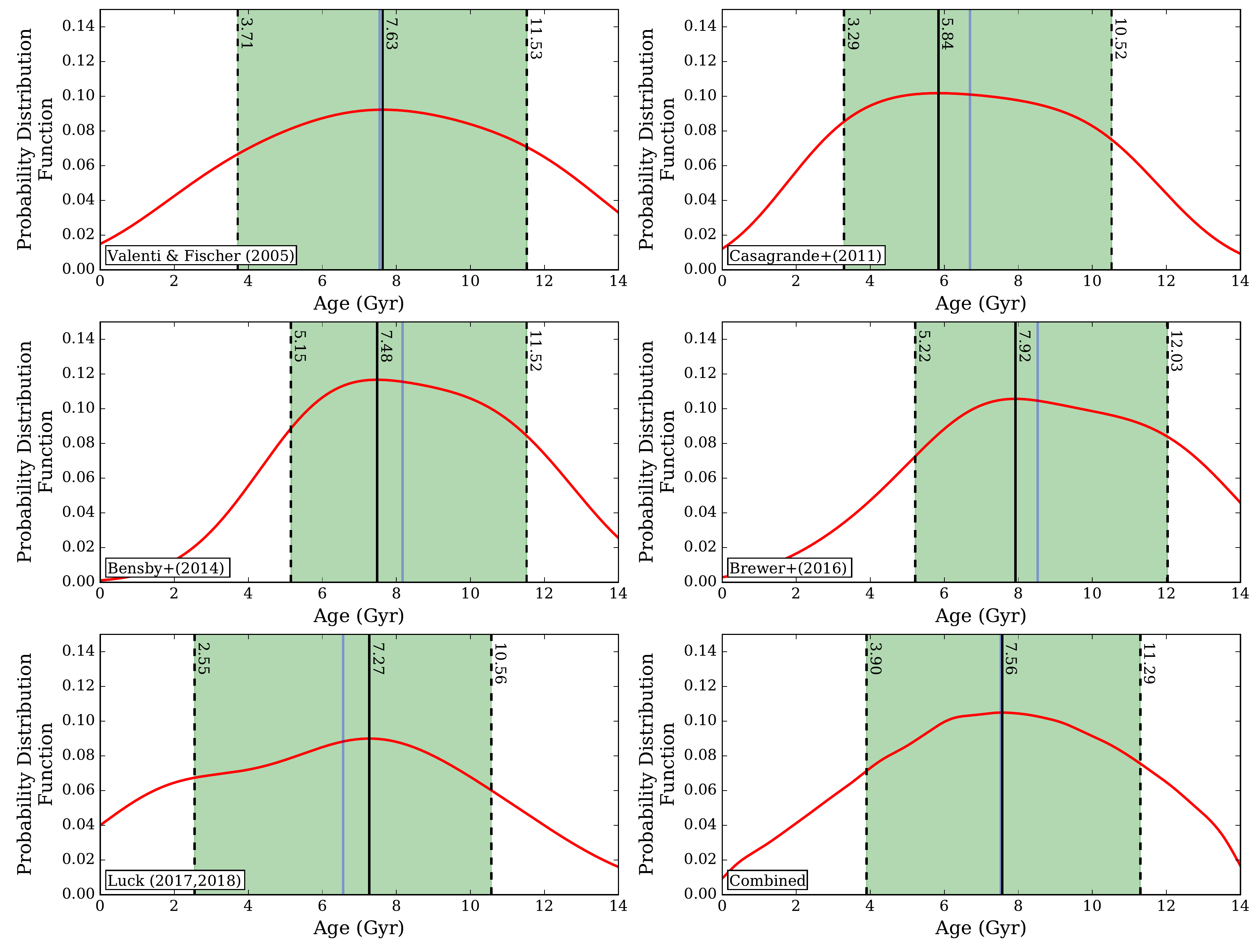}
\caption{Probability distribution functions for stars with $M \leq 1 M_{\odot}$, D $\leq$ 30 pc, and -0.52 $\leq$ [Fe/H] $\leq$ -0.28 from \cite{valenti2005}, \cite{casagrande2011}, \cite{bensby2014},  \cite{brewer2016}, and \cite{luck2017, luck2018}. The solid black lines indicate the maximum likelihood ages, while the dashed black lines indicate 16\% and 84\% probabilities. 50\% probabilities are indicated by light blue lines in all panels for comparison.  The bottom right panel shows the final combined probability distribution function from which the final metallicity-based age is taken.   }
\label{fig:feh}
\end{figure*}

Following the \cite{burgasser2017} approach, we select stars within 30 pc (using updated {\it Gaia} EDR3 astrometry) with $M \leq 1 M_{\odot}$ from all five samples. We limit each sample to have -0.52 $\leq$ [Fe/H] $\leq$ -0.28 based on our metallicity measurement for Ross 19A.  We assume a uniform probability distribution of age for each star between either the minimum and maximum ages provided (e.g., SPOCS) or between the 16\% and 84\% isochronal ages (e.g., GCS).  To ensure each sample of stars is given an equal weight, we calculate a probability distribution for each sample, which is then combined into a final probability distribution in a Monte Carlo fashion.  The probability distributions for each sample are shown in Figure \ref{fig:feh}. We find a maximum likelihood age of 7.56$^{+3.73}_{-3.66}$ Gyr, where the uncertainties reflect the 16\%--84\% probability ranges. 

Kinematics can also provide clues to age, as older objects have had more time to dynamically interact with their surroundings, resulting in higher overall velocities.  Using astrometry from {\it Gaia} EDR3 and the radial velocity for Ross 19A from \cite{jeffers2018} (-27.8$\pm$0.14 km s$^{-1}$), we find no clear membership in any young nearby moving group from BANYAN $\Sigma$ \citep{gagne2018}.  Following \cite{bensby2003}, we find the relative probability for thick disk to thin disk membership (TD/D) for Ross 19A to be 19\%.  

\begin{figure*}
\plotone{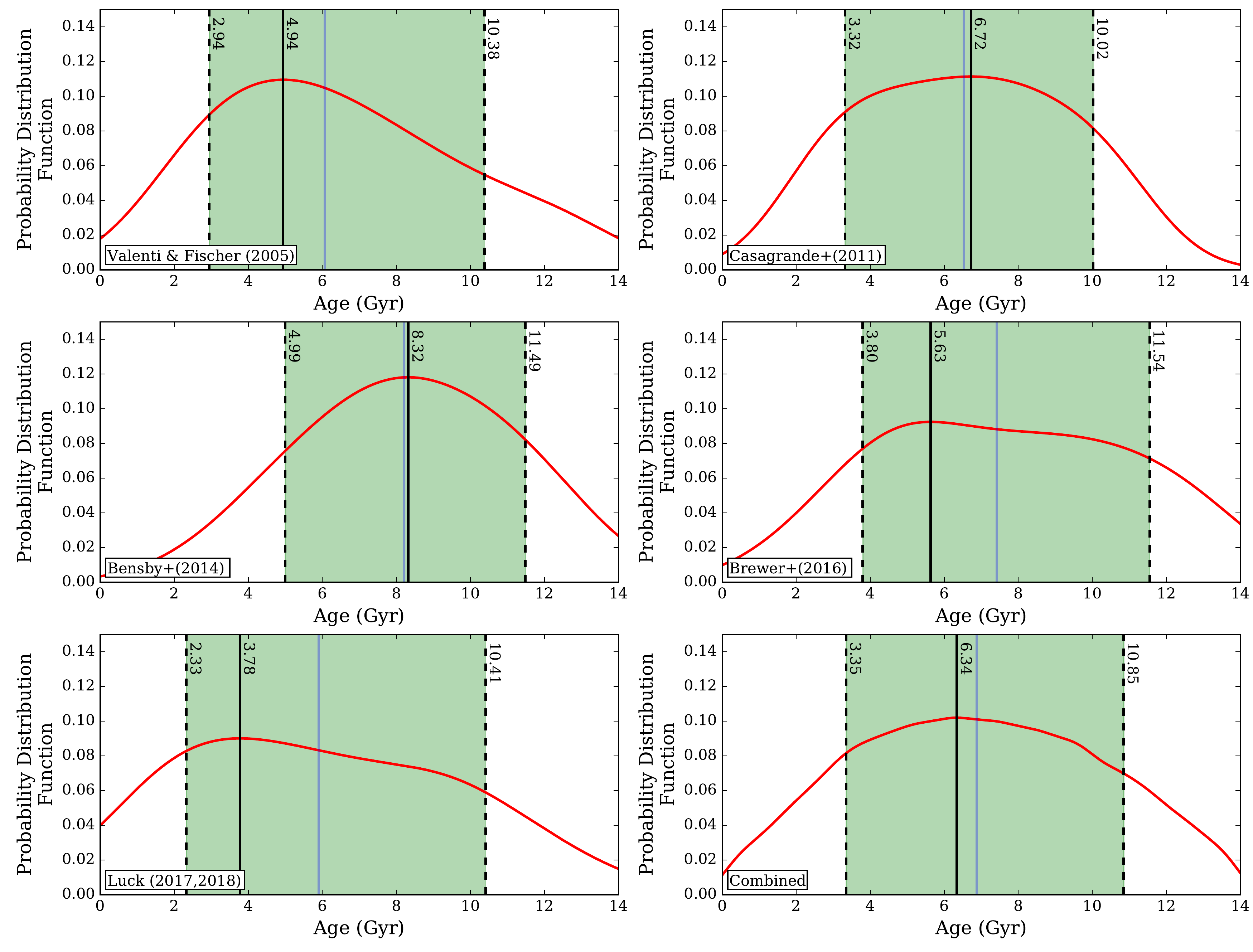}
\caption{Probability distribution functions for stars with $M \leq 1 M_{\odot}$, D $\leq$ 30 pc, and 56.4 km s$^{-1}$ $\leq$ ($U^2+V^2+W^2$)$^{1/2}$ $\leq$ 86.4 km s$^{-1}$ from \cite{valenti2005}, \cite{casagrande2011}, \cite{bensby2014},  \cite{brewer2016}, and \cite{luck2017, luck2018}. The solid black lines indicate the maximum likelihood ages, while the dashed black lines indicate 16\% and 84\% probabilities. 50\% probabilities are indicated by light blue lines in all panels for comparison.  The bottom right panel shows the final combined probability distribution function from which the final kinematic age is taken.}
\label{fig:vtot}
\end{figure*}

To derive a quantitative estimate of Ross 19A's kinematic age, we again use the \cite{valenti2005}, \cite{casagrande2011}, \cite{bensby2014}, \cite{brewer2016}, and \cite{luck2017, luck2018} samples to investigate age distribution as a function of total Galactic UVW velocity.  For each sample, we updated the astrometry for each star using {\it Gaia} EDR3.  We select stars from each sample with total UVW velocities within 15 km s$^{-1}$ of the total velocity of Ross 19A (71.43 km s$^{-1}$).  This range ensures at least 10 comparison stars from each sample. A larger bin size than $\pm$15 km s$^{-1}$ does not properly capture age gradients across velocity space.  For example, the transition in total velocity space between thin disk and thick disk stars happens over a range of $\sim$20 km s$^{-1}$ (e.g., \citealt{bensby2014}).  We find a maximum likelihood kinematic age of 6.34$^{+4.51}_{-2.99}$ Gyr, in agreement with our metallicity-based age calculation. The probability distributions for all samples are shown in Figure \ref{fig:vtot}.  

A summary of the age diagnostic information for Ross 19A is provided in Table \ref{tab:age}. We combine the age constraints from Ross 19A's metallicity and kinematics in a Monte Carlo fashion by randomly sampling the probability distribution functions from each and find a final age for Ross 19A of 7.2$^{+3.8}_{-3.6}$ Gyr. 

\begin{deluxetable}{lrccccccc}
\label{tab:age}
\tablecaption{Ross 19A Age Summary}
\tablehead{
\colhead{Property} & \colhead{Age Constraint}}
\startdata
Lithium & $\gtrsim$25 Myr\\
H$\alpha$ & $\gtrsim$600--800 Myr\\
NUV & $\gtrsim$600--800 Myr\\
Rotation & Unconstrained?\tablenotemark{a}\\
{[}Fe/H] & 7.56$^{+3.73}_{-3.66}$ Gyr\\
Kinematics & 6.34$^{+4.51}_{-2.99}$ Gyr\\
\colrule
Final Age & 7.2$^{+3.8}_{-3.6}$ Gyr\\
\enddata
\tablenotetext{a}{While we were not able to determine a rotation period for Ross 19A, low-amplitude variability and Ross 19A's small $v$sin$i$ measurement are consistent with an age $\geq$ several Gyr \citep{popinchalk2021}}
\end{deluxetable}

\section{Discussion}

\subsection{Binding Energy}
\label{sec:binding}
With a projected separation of $\sim$9900 au, the Ross 19AB system is one of a growing number of extremely wide low-mass companions (see e.g.,  \citealt{chinchilla2020, faherty2020}).  In order to compare to other known, wide, low-mass systems, we need mass estimates of both components of the Ross 19AB system.  For Ross 19A, we found a mass of 0.362$\pm$0.007 $M_{\odot}$ in Section \ref{sec:ross19a}.  

For Ross 19B, its spectral type range of T9.5$\pm$1.5 corresponds to an effective temperature range of 401--615 K \citep{kirkpatrick2020}.  Combined with our age constraint of 7.2$^{+3.8}_{-3.6}$ Gyr and the low-mass evolutionary models of \cite{phillips2020}, we find a mass range for Ross 19B of 0.015--0.038 $M_{\odot}$ (15--40 $M_{\rm Jup}$).  \cite{phillips2020} provides absolute magnitudes for their evolutionary models. If we also require $M_{\rm J}$ and $M_{\rm ch2}$ to be reasonably close (within 0.5 mag), we find a mass range of 0.017--0.024 $M_{\odot}$ (17--25 $M_{\rm Jup}$). 

We convert the projected separation of the system (568\arcsec\ at 17.444 pc) to a physical separation that accounts for inclination angle and eccentricity by multiplying by a factor of 1.16$^{+0.81}_{-0.32}$ following \cite{dupuy2011}.  Assuming Gaussian uncertainties, we find a physical separation of 11493$^{+8019}_{-3169}$ au.  Using our mass estimates for Ross 19A and Ross 19B combined with the separation above, we find a binding energy between 4.8$\times$10$^{+39}$ and 3.0$\times$10$^{+40}$ ergs.  Ross 19AB has a very low binding energy, rivaling those of young systems like Oph 11 \citep{close2007}, 2M1207 \citep{chauvin2004, chauvin2005} and COCONUTS-2b \citep{zhang2021b}, and near the minimum binding energy for substellar binaries (10$^{40}$ ergs; \citealt{close2007}).  We used the \cite{dhital2010} approximation of the \cite{weinberg1987} formulation for the dissipation lifetime for wide binaries, where the lifetime can be estimated by:

\begin{equation}
\tau = \frac{1.212 \times (M_1 + M_2)}{R} 
\end{equation}

\noindent where $\tau$ is the dissipation lifetime in Gyr, $M_1$ and $M_2$ are the masses of each component in Solar masses, and $R$ is the separation in pc.  For Ross 19AB, we find a dissipation timescale of 9.3--10.3 Gyr.  

An intriguing possibility is that Ross 19A is itself an unresolved binary.  If, for example, Ross 19A is an equal-mass binary, the binding energy of this system would double.  {\it Gaia} EDR3 includes several diagnostics that can be indicative of an object being non-single, most notably the Renormalised Unit Weight Error (\texttt{ruwe}).  Ross 19A has an \texttt{ruwe} value of 1.5, just above the value typically given for less reliable astrometric solutions (1.4), often used as an indicator for potential multiplicity.  Note, however, that other multiplicity diagnostics, such as the \texttt{ipd\_frac\_multi\_peak}, which is indicative of resolved close pairs, and the \texttt{ipd\_gof\_harmonic\_amplitude}, which indicates the level of asymmetry in {\it Gaia} images \citep{fabricius2021}, are consistent with Ross 19A being a single star.  The position of Ross 19A on color-magnitude diagrams also supports the single star hypothesis.  \cite{jeffers2018} investigated each of their targets for significant variations in their measured radial velocity values and found no evidence of spectoscopic binarity for this source. Further, Ross 19A has been imaged as part of the CARMENES High-resolution imaging survey \citep{cortes2017} and the Robo-AO M-dwarf Multiplicity survey \citep{lamman2020}, and found to be single in both surveys.  Both surveys place a similar limit on the presence of an equal-mass companion, with no companions found to separations $\gtrsim$0\farcs2  \citep{cortes2017, lamman2020}.    

There is also the possibility that Ross 19B is a close binary, as the binary frequency for very-wide low-mass companions is significantly higher than the field population (e.g., \citealt{faherty2010, law2010}). It has been suggested that extremely wide companions with separations greater than several thousand au need more mass to survive dynamically, and therefore very wide companions are often found to be multiples themselves. The mass ratio for this system is such that the binding energy would not change significantly if Ross 19B is found to be a multiple. However, it would be worthwhile to investigate whether or not Ross 19B is itself a binary via high-resolution imaging, considering its potential benchmark status. 

\subsection{Constraining Ross 19B's Formation Origin?}
Wide separation systems such as Ross 19AB give us the opportunity to explore formation and evolution mechanisms in a mass range where the coldest brown dwarfs and the largest giant exoplanets overlap. While core accretion and disk instability are currently the two primary mechanisms cited for the formation of giant planets in a disk, a third formation pathway exists for planetary-mass companions: turbulent fragmentation, a mechanism understood to produce binary stars (e.g., \citealt{offner2010, lee2017}).  The discovery of free floating brown dwarfs with masses $<$10 $M_{\rm Jup}$ (e.g., \citealt{liu2013, gagne2015, schneider2016, gagne2017, gagne2018b}) suggests that brown dwarf companions could form via core collapse with planetary masses.  

Recent studies (e.g., \citealt{mordasini2016, espinoza2017}) have shown that the composition of a gas giant exoplanet depends critically on where it formed within a protoplanetary disk.  Specifically, the C/O ratio of giant planets formed within a protoplanetary disk deviates significantly from that of their parent stars, where the total deviation depends on the specific formation location (e.g., inside or outside the water iceline).  Therefore, if Ross 19B  formed within Ross 19A's circumstellar disk and scattered via some interaction process (e.g., \citealt{malmberg2011, bromley2014}), their C/O ratios should be notably different, depending on formation location within the disk. Such a scattering event would require a third planetary or stellar component that would remain in close orbit around Ross 19A.  As of yet, there is is no evidence for such a companion (see Section \ref{sec:binding}).

Alternatively, if Ross 19B formed near its current location it would have formed well outside Ross 19A's protoplanetary disk, based on disk sizes of young M-type stars (e.g., \citealt{ansdell2018, long2018}). By measuring the atmospheric abundances of this cold companion (e.g., the C/O ratio), we may be able to shed light on its formation origin. Discrepant C/O ratios between Ross 19A and B may indicate significant scattering in Ross 19B's history, while consistent C/O ratios would provide new evidence of the often overlooked third pathway for giant planetary-mass companion formation. 

Retrieval techniques using low-resolution near-infrared spectra can extract C/O ratios for cold brown dwarfs at high-significance (e.g., \citealt{line2015, line2017, zalesky2019, gonzales2020}).  Ross 19B is so faint in the near-infrared that such a spectroscopic investigation will likely require the sensitivity of the {\it Hubble Space Telescope (HST)} or {\it James Webb Space Telescope (JWST)}. Other techniques are available for determining C and O inventories for low-mass stars.  For example, \cite{tsuji2014, tsuji2016} showed that a single high-resolution $K-$band spectrum can enable precise measurements of carbon and oxygen abundances of M-type stars.  Therefore Ross 19AB holds promise as an intriguing laboratory for which the origin of a companion close to the planetary mass boundary can be constrained.

\section{Summary}
We have presented the discovery of an extremely cold companion to the nearby M star Ross 19.  Based on new astrometry for this cold companion and a modified BANYAN~$\Sigma$ \citep{gagne2018} routine for co-moving companions, we find the likelihood that these two objects are a bound pair is 100\%.  We find a subsolar metallicity for Ross 19A, making Ross 19B one of a few low-metallicity, substellar benchmarks currently known.  As one of the widest and coldest known companions yet found, Ross 19B makes a compelling target for future spectroscopic characterization with {\it HST} or {\it JWST}.  Such observations may allow for the determination of the origin of this cold companion near the planetary mass boundary. Further astrometric observations could bring the small discrepancy between $\mu_{\delta}$ components of this system into better agreement and give an independent trigonometric parallax for Ross 19B. 

\acknowledgments
The Backyard Worlds: Planet 9 team would like to thank the many Zooniverse volunteers who have participated in this project, from providing feedback during the beta review stage to classifying flipbooks to contributing to the discussions on TALK. We would also like to thank the Zooniverse web development team for their work creating and maintaining the Zooniverse platform and the Project Builder tools. This research was supported by NASA grant 2017-ADAP17-0067. This material is based upon work supported by the National Science Foundation under Grant No. 2007068, 2009136, and 2009177. F.M.~also acknowledges support from grant 80NSSC20K0452 under the NASA Astrophysics Data Analysis Program.  E.G. acknowledges support from the Heising-Simons Foundation.  (Some of) The data presented herein were obtained at the W. M. Keck Observatory, which is operated as a scientific partnership among the California Institute of Technology, the University of California and the National Aeronautics and Space Administration. The Observatory was made possible by the generous financial support of the W. M. Keck Foundation.  This publication makes use of data products from the {\it Wide-field Infrared Survey Explorer}, which is a joint project of the University of California, Los Angeles, and the Jet Propulsion Laboratory/California Institute of Technology, and NEOWISE which is a project of the Jet Propulsion Laboratory/California Institute of Technology. {\it WISE} and NEOWISE are funded by the National Aeronautics and Space Administration.  Part of this research was carried out at the Jet Propulsion Laboratory, California Institute of Technology, under a contract with the National Aeronautics and Space Administration.  The authors wish to recognize and acknowledge the very significant cultural role and reverence that the summit of Maunakea has always had within the indigenous Hawaiian community.  We are most fortunate to have the opportunity to conduct observations from this mountain.

\software{\texttt{crowdsource} \citep{schlafly2018, schlafly2019}, BANYAN~$\Sigma$ \citep{gagne2018}, CoMover \citep{gagne2021}; SEDkit \citep{filip2020}, SpeXTool \citep{cushing2004}}

\end{document}